# Methods of Stellar Space-Density Analyses: A Retrospective Review

*B. Cameron Reed, Department of Physics (Emeritus), Alma College*
*Alma, MI 48801   reed@alma.edu*

June 17, 2020

*Stellar space density analyses, once a very active area of astronomical research, involved transforming counts of stars to given limiting magnitudes in selected areas of the sky into graphs of the number of stars per cubic parsec as a function of distance. Several methods of computing the transformation have been published, varying from manual spreadsheets to computer programs based on very sophisticated mathematical techniques for deconvolving integral equations. This paper revisits these techniques and compares the performance of seven of them published between 1913 and 2003 as applied to both simulated and real data.*

## 1.   Introduction

The origins of quantitative efforts to analyze the distribution of stars within the Galaxy date to the late eighteenth and early nineteenth centuries with the work of Sir William Herschel (1738-1822), who attempted to determine what he called the "construction of the heavens" by "star gauging". By counting the numbers of stars to successively fainter magnitude limits in some 700 regions of the sky and assuming that all stars were of the same absolute magnitude, Herschel was able to deduce the relative dimensions of the Galaxy, concluding that the Sun was near the center of a flattened, roughly elliptical system extending about five times further in the plane of the Milky Way than in the perpendicular direction. This work was carried on by his son, Sir John Herschel (1792-1871), who used stellar parallaxes to establish the first evidence of an intrinsic distribution of stellar absolute magnitudes: the Luminosity Function (LF). German astronomer Hugo von Seeliger (1849-1924) put stellar distribution analyses on a firm mathematical footing by establishing integral equations relating the run of density of stars as a function of parallax (distance) to the number appearing in a given interval of apparent magnitude, as well as procedures for correcting the apparent distribution of stars for the effects of interstellar extinction.

What might be termed the "classical golden age" of stellar distribution analyses reached its zenith in the first quarter of the twentieth century with the work of Jacobus Kapteyn (1851-1922) at the University of Groningen. Kapteyn initiated a plan to obtain star counts, magnitude estimates, spectral classifications, proper motions and radial velocities for about 20,000 stars in some 200 "Selected Areas" distributed around the celestial sphere. The resulting picture of galactic structure, the "Kapteyn Universe", was published in *The Astrophysical Journal* shortly before his death (Kapteyn 1922), and depicted a Sun-centered lens-shaped structure about 16 kiloprasecs in diameter with an axial-to-polar ratio of about 5:1. Kapteyn noted that the system could not be in a steady state unless there was a general rotational motion around the galactic polar axis, and also that when the work was further developed, it might be possible to "... determine the amount of dark matter from its gravitational effect." While he surely had in mind ordinary but non-luminous matter as opposed to current particle-physics concepts of dark matter, his remark was prescient.

The subsequent discoveries of interstellar extinction, galactic rotation, and different stellar populations radically altered astronomers' ideas of galactic structure, but Herschellian star gauging continued to be of value. Space-density analyses entered a second golden age in the post-World-War-II period, in particular through the efforts of American astronomer Sidney W. McCuskey (1907-79) and his collaborators and students at the Case Institute of Technology in Cleveland, Ohio (later and still Case Western Reserve University). McCuskey undertook, from 1945 onwards, an extensive series of stellar-distribution analyses with the goal of determining the galactic structure within 1500 parsecs of the Sun. In his last publication, which appeared posthumously,

McCuskey analyzed data for over 8,000 stars in a field of sky in the Centaurus-Crux region of the southern sky (McCuskey 1983).

The raw data for these studies were apparent magnitudes and spectral classifications for hundreds or thousands of stars in some target area of the sky which usually covered a few or tens of square degrees. These were usually acquired by calibrating and examining photographic plates obtained with wide-field Schmidt telescopes. The stars would be segregated into relatively narrowly-defined groupings of spectral types (for example, B8-A0 dwarfs; the reason for this is described in more detail below), and counts made of the number of stars which appeared in given ranges of apparent blue ($B$) or visual ($V$) magnitude. For practical purposes, the objective-prism spectra obtained in such studies could be reliably classified to $V \sim 13$.

Turning these data into runs of the stellar space density required two steps. First, and the main focus of this paper, it was necessary to "deconvolve" the star counts to produce a so-called "fictitious" density function, the run of density versus distance neglecting any interstellar extinction. The second step was to correct the fictitious densities for the effects of extinction by using information on stellar colors to determine the run of extinction versus distance; this was usually done by studying intrinsically-bright O and B-type stars in the field of interest which could be detected to great distances.

As might be imagined, both steps were fraught with statistical uncertainties. At bright and faint magnitudes, the counts could be small in number due to a sheer lack of nearby stars and running up against the effects of detectability limits. Whatever uncertainties were baked into the fictitious densities would then be compounded in modeling the run of extinction. There was by no means any guarantee that extinction would be a smooth function of distance: Nature might be so insensitive to the needs of astronomers as to distribute interstellar clouds and dust in globules and sheets between the Sun and the stars under study.

My purpose in this paper is to examine the relative merits of various methods that were developed to determine the run of fictitious density; I do not concern myself with the issue of extinction corrections for the reasons adduced above.

The fundamental difficulty in determining the fictitious densities is that since the star counts are related to the density by an integral equation [see Eq. (2) below] with the desired density function residing *within the integrand*, the density function $\rho(r)$ is not uniquely related to the number of stars as a function of apparent magnitude. This situation renders multiple solutions possible, with no infallible way of telling which is the "true" one. Various approaches to recovering the density distribution were developed, often reflective of the computational technology of their time.

The outline of this paper is as follows. In Section 2 I give a brief summary of the fundamental integral equation of stellar statistics that incorporates the desired density function, and hand-computation methods of solving it. Section 3 summarizes seven solution methods that have been developed over nearly a century. To compare these as uniformly as possible, I have developed programs for generating simulated star counts corresponding to assumed density and luminosity functions; this is described in Section 4. Results of the comparisons are presented in Section 5, and Section 6 applies what appears to be the best methods to two examples of real data. Some brief closing remarks appear in Section 7.

## 2. The Integral Equation of Stellar Statistics, (*m*, *log π*) Tables, and The Revival of Star Counts

As described above, let $\rho(r)$ represent the number of stars per cubic parsec of the restricted range of spectral types under consideration as a function of fictitious distance *r* in some direction of interest. A thin spherical shell of space of thickness *dr* at distance *r* will then contain a total of $4\pi \rho(r) r^2 dr$ stars. For practical purposes one will be able to image only so many square degrees of sky, usually designated as $\Omega$. The total number of square degrees on the celestial sphere is 41253, so this reduces the count in a distance shell to $(4\pi \Omega/41253)\rho(r)r^2 dr$ stars. At this point the *Luminosity Function* (LF) is introduced, usually designated as $\Phi(M)\,dM$, which gives the fraction of stars of the type under consideration which have absolute magnitudes *M* between *M* and *M* + *dM*. With this, the number of stars in the shell that will appear to have *apparent* magnitudes between *m* and *m* + *dm* will be $(4\pi \Omega/41253)[\Phi(M)dM]\rho(r)r^2 dr$, where *M*, *r*, and *m* are linked by the usual distance-modulus equation,

$$m - M = 5(\log r) - 5. \tag{1}$$

Formally, $\Phi(M)$ should be written as $\Phi(M, r)$.

Integrating over all space with *M*, *r*, and *m* related in this way at every distance gives the total number of stars that will appear to have *apparent* magnitudes between *m* and *m* + *dm*, usually designated as *A*(*m*):

$$A(m) = \frac{4\pi \Omega(\Delta m)}{41253} \int_0^\infty \Phi(M)\rho(r)r^2\,dr, \tag{2}$$

where *dM* has been written as *Δm* in recognition of the fact that for practical purposes the ranges of apparent magnitudes will be finite; typically, $\Delta m \sim 0.5$ mag. Equation (2) is known as the Fundamental Integral Equation of Stellar Statistics.

The reason that a restricted range of spectral-luminosity types is considered is to narrow the range of absolute magnitudes that will dominantly contribute to $\Phi(M)$, and thus hopefully return sharper resolution in determining $\rho(r)$. In most restricted-type studies the LF was usually assumed to be Gaussian with a mean absolute magnitude $M_O$ and dispersion $\sigma$:

$$\Phi(M) = \frac{1}{\sigma\sqrt{2\pi}} exp\left[-\frac{1}{2\sigma^2}(M - M_O)^2\right]. \qquad (3)$$

Equation (2) is an example of what is known to mathematicians as a convolution problem: the function to be determined, $\rho(r)$, lies within the integrand and is *convolved* with $\Phi(M)$. While mathematicians have been studying integral equations for well over a century and have devised numerous ways of addressing them, the fundamental stumbling block is that the problem is ill posed: Different possible solutions for $\rho(r)$ can sensibly reproduce the stars counts (say, within their Poisson errors), so there is no unique solution for $\rho(r)$. Since star-count data can be plagued by small-number statistics which carry inherently large relative errors, the situation becomes even more tangled.

As an example of the problems that can arise in determining $\rho(r)$, consider the following very straightforward approach to solving equation (2). Suppose that the star counts are specified for *K* values of *m*, that is, one has $A(m_j)$ with *j* = 1 to *K*. Treat the counts as a one-column matrix $[A(m_j)]$ of *K* rows. Divide space up into some sensible number of shells of distance $r_i$, with *i* = 1 to *N*, and consider the values of the stellar density at each distance, $\rho(r_i)$ as a one-column matrix of *N* rows, $[\rho(r_i)\, r_i^2]$, whose elements incorporate the $(4\pi\Omega\,\Delta m/41253)$ prefactor in equation (2). Then construct a matrix of dimension *N* rows by *K* columns, $[\Phi_{ij}]$, whose elements are computed from the luminosity function with $M = m_j - 5(\log r_i) + 5$. Invert $[\Phi]$, multiply it into $[A(m_j)]$, and *voila*: out pops $[\rho(r_i)]$. The problem that inevitably arises with this approach, however, is that the results often include physically absurd negative densities. Least-squares-type minimizations of the differences between predicted and observed star counts also tend to generate negative densities. The underlying problem is that the mathematics is ignorant of the physics and does not know that negative densities are impossible.

Analytic solutions to extracting $\rho(r)$ were developed early on (see below), but their inflexibility drove early researchers to a labor-intensive manual-spreadsheet method known as an (*m*, *log π*) table. This approach effectively mimicked the matrix approach, but with a human being at the helm whose job was to ensure realistic densities. In the spreadsheet, rows correspond to apparent magnitude and columns to shells of volume set out in increments of equal spacings of (*log r*) which ran over a range of distances sensible to the data at hand. The "*log* $\pi$" nomenclature arose from the fact that stellar distances could be determined via trigonometric parallaxes, whose values were always designated by $\pi$. The user would first calculate a matrix of values of *Φ* according as the apparent magnitudes and distances corresponding to the elements of the spreadsheet. Then an initial guess as to the density distribution would be made, say by assuming a uniform density over all distances. Predicted star counts would be generated and compared to those observed, and a lengthy iterative process of adjusting the density distribution would then follow until the predicted and observed counts were in reasonable agreement. Details of this procedure are laid out in classic texts by Bok (1937), Trumpler and Weaver (1953), Mihalas and Routly (1968), and Mihalas and Binney (1981), but, as might be imagined, there was no small amount of subjectivity involved.

From the 1960's through the 1980's, approaches to star-count analyses underwent two major revisions. First, the arrival of mainframe and later personal computers meant that humans could be freed from the drudgery of manual calculations, and several algorithms for solving equation (2) were advanced. As described in the following section, some of these reached levels of mathematical sophistication that were probably unjustified in view of the nature of the input data. This author admits to developing one such technique (Reed 1983), an indiscretion I now attribute to the youthful misconception that a closed-form mathematical solution to a physical problem must always be preferred. The second revision saw the emergence of an entirely new philosophy of the role of star counts. This was that instead of using measured counts with all their statistical noisiness to try to work backwards to $\rho(r)$, approach the problem from the other direction by *assuming* a model of the galaxy which comprised various components (thin disk, thick disk, halo, ...), *predict* star counts, and then adjust the model as necessary. This change in was prompted by evolutions in technology which brought forward high-speed plate scanning machines, efficient large-area CCD detectors, the troves of data that could be acquired with ever-larger ground- and space-based telescopes which could reach very faint limiting magnitudes, and which have culminated in the *HIPPARCOS* and *Gaia* missions that measured parallaxes directly. As examples of such models, see Bahcall & Soneira (1980) and Pritchet (1983) and references therein. Bahcall (1986) remarked that star counts " ... were a subject whose time had passed and come again." With these developments, the classical (*m*, *log* $\pi$)-type approach rapidly faded into history.

While describing some of this history to a group of students, I began to wonder if the various density-recovery methods that had been developed had ever been directly tested head-to-head by feeding them simulated data derived from an assumed-known density distribution. A search of the literature turned up seven methods which could readily be programmed and compared, including two published by myself. While some papers reported comparisons between methods, none ever undertook a comparison of all of them with the same data. As described in the following section, two other methods have also been published, but I disregard them for various reasons.

Before proceeding to describe these methods, another remark on the nature of density-recovery techniques is appropriate. Suppose that we are working with a Gaussian LF as in equation (3); this is in fact assumed to be the case in generating the simulated data in Section 4 below. There is a 50% chance that a given star will have an absolute magnitude brighter (fainter) than the mean absolute magnitude $M_O$. For a given apparent magnitude, this means a 50% chance that the star will be inferred to be more (less) distant than the nominal distance $r_O$ corresponding to $M_O$. However, the impact of this on the density distribution is asymmetric. If the star is imagined to be closer than $r_O$ it will lie in a smaller volume of space than if it is more distant, which causes the inferred stellar density at nearby distances to be greater than is actually the case. Conversely, the range of distances available for fainter magnitudes (distances greater than $r_O$) is unlimited, with the result that inferred density distributions will inevitably exhibit long low-density exponential tails. These effects cause the distance distribution for an individual star to have a mean distance value $\langle r \rangle$ greater than $r_O$, namely $\langle r \rangle = r_O e^{\lambda}$, where $\lambda = (\sigma \ln 10)^2 / 50$, while having a most probable distance of $r_{mp} = r_O e^{-2\lambda}$ and making a maximum contribution to the run of density at $r_{max} = r_O e^{-6\lambda}$. For values of $\sigma$ normally adopted in density analyses, these values do not differ wildly from $r_O$ (for example, $r_{max} \sim 0.85 r_O$ for $\sigma = 0.5$), but do cause the distance distribution inferred for all stars to be asymmetric. These "front-loading" and "exponential tail" asymmetries are built in to every stellar density analysis, but were often not explicitly acknowledged in practice. These effects may have contributed to McCuskey's (1965) observation in a review article on galactic structure that " ... there is evidence from the general star counts for a somewhat elongated relatively high density region near the sun with its maximum about 300-500 parsecs from the sun and toward the galactic anti-center."

## 3. Stellar Density Analysis Methods

The seven methods alluded to above are summarized here, in chronological order of publication.

### 3.1 Eddington (1913)

Eddington (1913) developed a direct solution to equation (2) by a Taylor-series expansion method. His notation is somewhat confusing; good summaries of his approach appear in Spaenhauer (1977) and Ochsenbein (1980).

This method assumes a Gaussian LF of dispersion $\sigma$. If $A(x)$ is the number of stars with distance modulus $m - M$ between $x$ and $x + dx$, then the true number of stars $T(x)$ with distance moduli between these limits is given by

$$T(x) = A(x) - \frac{\sigma^2}{2}\left(\frac{d^2A}{dx^2}\right)_x + \ldots. \tag{4}$$

The second derivatives of the star counts can readily be computed via the differences of successive counts,

$$\left(\frac{d^2A}{dm^2}\right)_m = \frac{1}{\Delta m^2}\left[A(m+\Delta m) + A(m-\Delta m) - 2A(m)\right]. \tag{5}$$

This method is very amenable to spreadsheet computation. Star counts as a function of magnitude are laid out in the rows of the spreadsheet. The $x$ value of each row is determined from $x = m - M_O$, and distances corresponding to the inner and outer limits of each magnitude bin are computed from $x \pm \Delta m/2$. Each magnitude bin thus corresponds to a spherical shell whose volume can be computed on accounting for the areal coverage $\Omega$ involved, and values of $T(x)$ can be directly converted into densities for each bin. In addition to assuming a Gaussian LF, the limitations of this method are that the second derivatives may be erratic, and that densities can be computed only at distances corresponding to the centers of the count bins.

3.2     Crowder (1959)

This method seems to have been the first explicitly developed for its adaptability to machine computation. If it is assumed that the density distribution follows the description

$$\rho(r) = \exp\left[a + b(\ln r) + c(\ln r)^2\right], \tag{6}$$

then equation (2) can be integrated directly in the case of a Gaussian LF with the result that the star counts will behave as

$$\ln A(m) = C_1 + C_2 m + C_3 m^2. \tag{7}$$

The coefficients $(a, b, c)$ can be determined in terms of $(C_1, C_2, C_3)$; a parabolic fit to $\ln A(m)$ then automatically determines the run of density. This approach was used extensively in the Case/McCuskey studies mentioned above. While it has the advantage of computational convenience, there is obviously no *a priori* guarantee that the density profile will behave as equation (6); indeed, such a prescription can at most accommodate only a single maximum in $\rho(r)$, at $ln\ r = -b/2c$. With the curve-fitting routines built into modern spreadsheets, this method is also easily programmed on a desktop computer.

### 3.3   Dolan (1974)

This method was also developed at Case Western Reserve University, and involves an approach that derived from computations involved in analyses of astronomical X-ray spectra. The LF is again assumed to be Gaussian. Dolan's notation and description can be confusing in places, but he does give a detailed example against which users can check their own implementations. This method, which is also very easy to set up on a spreadsheet, includes a "smearing" matrix which corresponds to the spread of the LF and which in effect leads to a sort of running-average of counts over distance bins in a manner similar to Eddington's second-derivative technique. In practice, I have found that the two methods produce very similar results.

### 3.4   Gschwind (1975)

This method is very computationally intensive, but requires no specific form for the LF. For each star in the sample, a random absolute magnitude is generated, consistent with the LF. With each stellar distance then (presumed) known, the density distribution is computed. This "Monte Carlo" process is then repeated numerous times (Gschwind suggested 100 trials), allowing one to determine a mean density distribution and estimate the associated uncertainty.

### 3.5   Reed (1983)

This technique also assumes a Gaussian LF and involves a direct analytic solution to equation (2). First, the common logarithms of the star counts are fit by a linear function,

$$log\ A(m) = s_0 + s_1 m. \qquad (8)$$

The fit need not be particularly good; the only purpose of this step is to determine the coefficients $s_0$ and $s_1$. The star counts are then reduced to be all on the order of unity by the transformation

$$A^*(m) = A(m)\ 10^{-(s_0+s_1 m)}. \qquad (9)$$

These reduced counts are then fit with a polynomial of order up to a few. With the star counts in this form, equation (2) can be transformed into a form where the density function emerges as a sum of Hermite polynomials whose expansion coefficients are related to the coefficients of the polynomial fit to the $A*(m)$.

When I first developed this method, I tested it against some simulated data, achieving positive results. While my memory is now vague as to exactly how I generated the simulated data, I do not now believe that this was a rigorous test. The data were likely simulated counts generated by a numerical integration scheme as opposed to generating a list of individual simulated apparent magnitudes and then binning; the method likely essentially reproduced what was input to it. More rigorous – and unfortunately less encouraging – tests are described in the following section.

3.6     Reed (1985)

This method is a computerized version of the traditional hand-computed ($m$, $log \pi$) table that is not only much faster than a manual computation but removes most of the subjectivity that can be introduced by a human operator. The key to this method is a very general, intuitively-appealing iterative approach to solving integral equations published by Lucy (1974), which has been cited over 1,500 times. A great advantage of this approach is that if the user's initial guess for the density distribution is positively-valued, all subsequent estimates will be as well. At each step, predicted star counts are compared to observed ones, and the density distribution is refined to bring the two into closer alignment. The user selects both the maximum number of iterations allowed, and, as in a traditional ($m$, $log \pi$) table, the stepsize in ($log r$). From experimenting with this method over many years I have found that only a few iterations (generally ~ 10) are required to bring as many of the predicted counts within the Poisson errors of the actual counts as can be sensibly had without introducing wild oscillatory variations in the density distribution that are unlikely to be physically realistic. For $\Delta(log r)$, a value on the order of 0.05 usually gives a sensible number of volume shells. No specific LF is assumed.

3.7     Branham (2003)

This matrix-based approach utilizes the mathematics of Regularized Total Least Squares, which is particularly effective for addressing systems of ill-conditioned linear equations. No specific LF is assumed. Like the Reed (1985) method above, this method is iterative, with the user adjusting a quantity known as the "ridge parameter" (designated as $\tau$), generally adjusting it upward until no negative values of the density appear and the density distribution looks reasonable. (If negative densities do appear, they should be tolerated only at the brightest and faintest magnitude bins, where the relative errors in the counts are large). For reasons explained by Branham, the ridge parameter can take quite large values. The matrix mathematics underlying this method are involved, but Branham presents a very explicit recipe for implementing the method. In view of the fact that the sizes of the matrices involved will vary from case to case, this method is best implemented with a program where these can readily be varied, as opposed to using a spreadsheet.

Two other published methods are not used here: Spaenhauer (1978) and Ochsenbein (1980). Spaenhauer developed a least-squares matrix approach, but concluded on the basis of test data that " ... when we do not know a prior the location of a density maximum [spiral arm], it is very doubtful whether the star numbers obtained ... are reliable." In the case of Ochsenbein, I must confess to finding his mathematics to be impenetrable; his method does not seem to have been used in practice.

In comparing these methods in what follows, I do not concern myself with comparing their approaches to computing uncertainties in the density estimates (which can be very mathematically elaborate), or with more subtle effects such as the Malmquist bias.

4. **Simulated Data**

To test the above routines, I have generated simulated data corresponding to two assumed density distributions. To generate this data, I wrote a master program in which the desired density distribution $\rho(r)$ is programmed into a subroutine. The LF is assumed to be Gaussian, with values of $M_O$ and $\sigma$ specified by the user; this could be altered if desired. The user also specifies the simulated survey area $\Omega$ and the distance limit $r_{max}$ to which calculations are to be carried out.

The program begins by first numerically integrating $\rho(r)$ out to the specified distance limit to determine the number of stars $N$ to simulate; this is done with a 2,000-slice Simpson's rule subroutine. This subroutine also determines the maximum value of $r^2\rho(r)$; call this $\left[r^2\rho(r)\right]_{max}$. Generating a distribution of distances that respects the density distribution is done within another loop which, upon each execution, calls the computer's built-in random-number generator to create a pair of random numbers, $(x, y)$, both in the range (0, 1). $x$ is scaled to be in the range (0, $r_{max}$) to create a trial distance $r_{trial}$. The value of $r_{trial}^2\rho(r_{trial})$ is computed, and if the "paired" $y$ lies between 0 and $r_{trial}^2\rho(r_{trial})/\left[r^2\rho(r)\right]_{max}$, then the distance is taken as valid. If the random distance is deemed invalid, the loop executes again until a valid one is found. This process is repeated as many times as necessary until distances for all $N$ stars have been generated.

This generates a list of distances consistent with the assumed density distribution since the number of stars at any distance will be proportional to $r^2\rho(r)$. A similar loop then generates as many absolute magnitudes according as the specified LF by sampling absolute magnitudes within $\pm 3$ standard deviations of $M_O$. Each absolute magnitude is assigned to a star, and corresponding apparent magnitudes are computed. A file of all data is produced, and a separate program bins the star counts according as bin centers and widths specified by the user. Since a finite number of stars are generated, there will naturally be some noise in sampling the assumed density distribution; only if an effectively infinite number of stars were used would the distribution be smoothly sampled.

The first assumed distribution is a simple one, but likely to prove a challenge for any deconvolution method: no stars out to a distance of 100 pc, then a uniform density of one star per 100 pc$^3$ out to 500 pc, and none at greater distances. I assumed $M_O = 0$, $\sigma = 0.5$, and $\Omega = 20$ square degrees, which give $N = 2520$, a largish but not unreasonable sample for a traditional analysis.

The second assumed distribution has an exponentially-decreasing "background" density punctuated by a Gaussian-shaped "spiral arm" at a specified distance. The functional form is

$$\rho(r) = Ae^{-(r-r_0)/B} + Ce^{-(r-r_1)^2/D}, \tag{10}$$

where $A$, $B$, $C$, $D$, $r_0$ and $r_1$ are set by the user. In the test described below I adopted ($A$, $B$, $C$, $D$, $r_0$, $r_1$) = (0.05 stars/pc$^3$, 300 pc, 0.03 stars/pc$^3$, 10$^4$ pc$^2$, 0 pc, 500 pc) with a limiting distance of 1,000 pc. With $M_O = 5$ (similar to that of the Sun), $\sigma = 0.5$, and $\Omega = 5$ square degrees, $N$ emerges as 4726 stars between magnitudes 6.4 and 16.3. This is a large number for a traditional analysis, but should give the routines enough data to converge on stable solutions.

5. **Comparing the Methods – Simulated Data**

Spreadsheets or FORTRAN programs were written for each of the above methods: the former for the Eddington, Crowder, Dolan, and Reed (1983) approaches, and the latter for the others. The author would be happy to distribute copies of the spreadsheets and codes to any interested reader.

For the uniform-density model, the star counts were binned into thirteen half-magnitude intervals centered on $m$ = 4.0 (0.5) 10.0. For the Reed (1983) Hermite-polynomial method, a sixth-order fit was performed to the reduced star counts. For the Gschwind method, 1000 Monte-Carlo trials were performed, and for Branham's method a ridge parameter value of $10^6$ was used; this resulted in only some very slight negative densities beyond distances of 600 pc. For the Reed (1985) automated routine, $\Delta (log\ r) = 0.05$ was adopted.

As can be seen in Figure 1, the Crowder, Gschwind, and Hermite-polynomial methods all produce terrible results, exhibiting serious front-loading and long exponential tails.

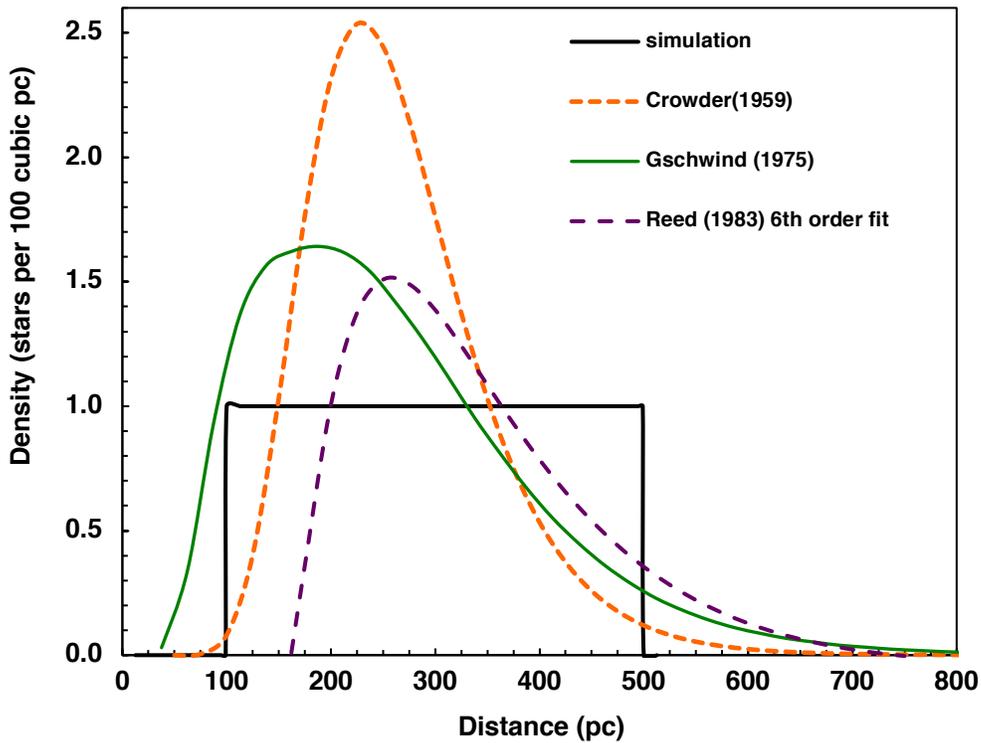

Figure 1. Results of star-count analyses for uniform-density model. See also Figure 2.

The Eddington, Dolan, Reed (1985), and Branham (2003) methods performed more respectably (Figure 2), with the Eddington and Dolan methods generating practically identical results.

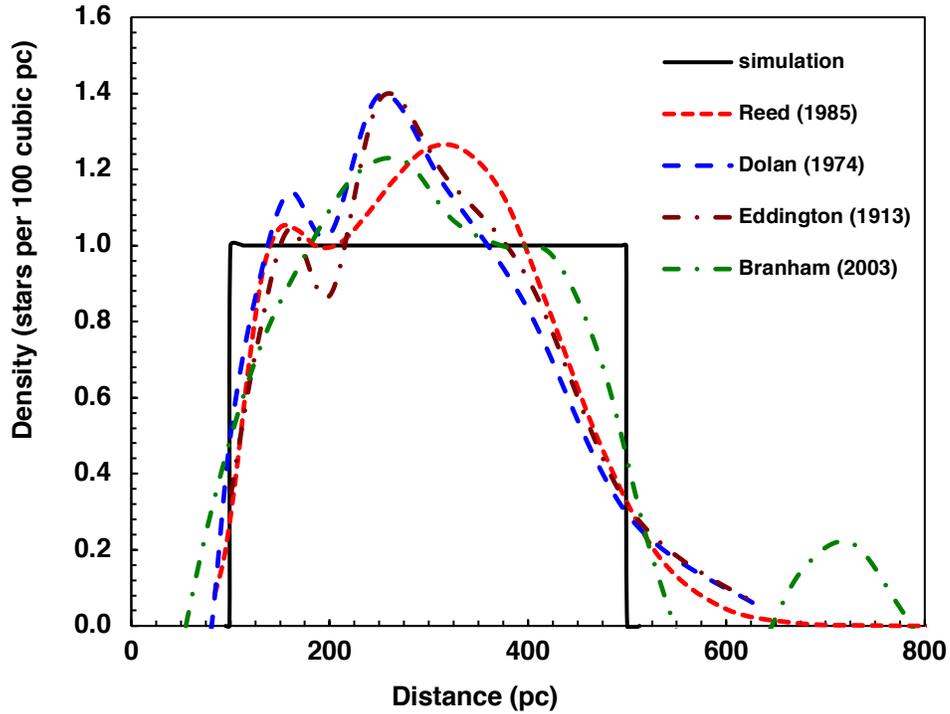

Figure 2. Results of star-count analyses for uniform-density model.

In the case of the dual-exponential model, the situation is the same: Crowder, Gschwind, and Reed (1983) turn in abysmal performances, while Eddington, Dolan, Reed (1985) and Branham do respectable jobs of capturing the Gaussian maximum in the density at 500 pc, if one disregards oscillations at nearby distances (Figure 3). Branham captures the location and value of the maximum most closely, but puts secondary peaks at about 275 and 800 pc; these are not likely to be of great statistical significance. The value of the ridge parameter here is $\tau = 70{,}000$, essentially the lowest value which gave no negative densities. Reed, Dolan and Eddington all place the peak density at about 450 pc (not far from the $r_{max}$ position described above), with Reed's density run being the smoothest of the three: 15 iterations gave 16 of 21 star-count values reproduced within their Poisson errors. All three of these latter methods, underestimate the maximum density by about 20%, a manifestation of the front-loading/exponential-tail phenomena.

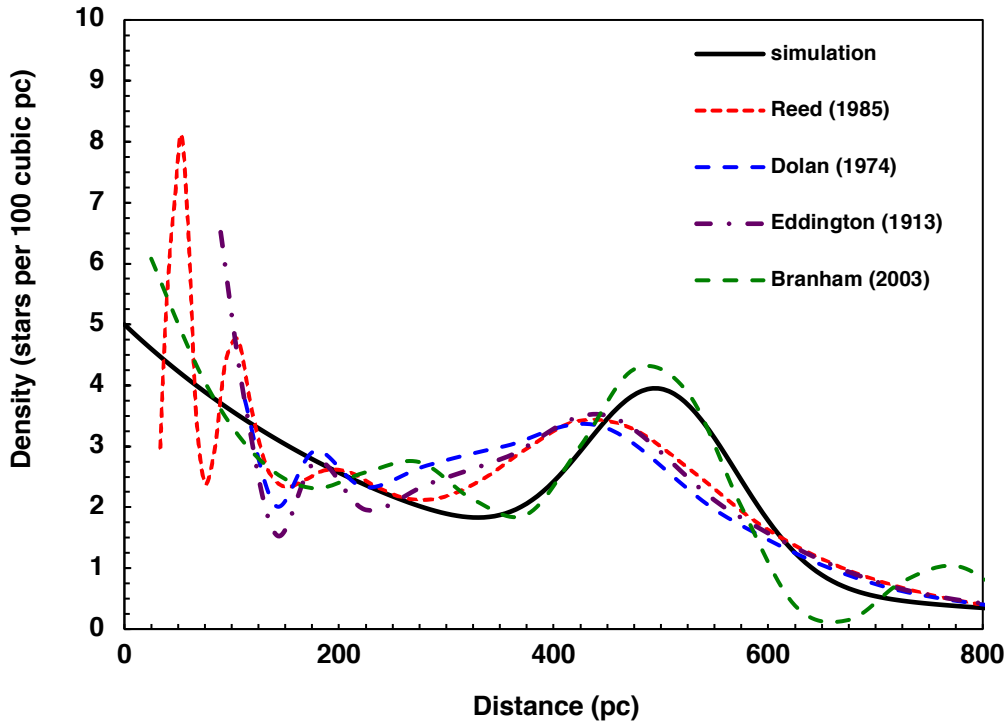

Figure 3. Results of star-count analyses for the double-exponential model of Eq. (10).

Based on these tests, Branham's method seems to have the edge, with the old-school Eddington and automated ($m$, $\log \pi$) methods not far behind. In both Eddington's and Dolan's methods, it is not clear from their papers how non-Gaussian luminosity functions would be accommodated.

## 6. Comparing the Methods – Real Data

### 6.1 South Galactic Pole M-Dwarf Data

In a paper published shortly after Dolan's (1974) paper, Thé and Staller (1974) examined the distribution of M-type dwarfs in the direction of the south galactic pole. Their overall sample comprised 96 M2-M4 dwarfs in a 238-sqaure degree region with photographic magnitudes between about 12.5 and 16. Upon eliminating a few extremely faint stars of uncertain spectral type and a few high-velocity stars, they were left with 82 objects on which they performed a traditional ($m$, $\log \pi$) analysis (apparently by hand), assuming an average absolute magnitude $M_{pg} = 12.25$ and dispersion $\sigma = 0.5$ mag. Since all of these stars are nearby, no corrections for interstellar extinction were applied. Their analysis revealed a maximum density of about four stars per 100 pc$^3$ at $r \sim 20$ pc, and an average density of about three stars per 100 pc$^3$ beyond this distance. On accounting for various selection effects, they estimated that the true average density would be closer to about six stars per 100 pc$^3$. This paper garnered attention at the time because their average density was significantly lower than that found by Murray and Sanduleak (1972) for the same type of stars in the direction of the north galactic pole, about 12 stars per 100 pc$^3$.

Dolan (1975) re-analyzed the Thé and Staller data (83 stars) with his technique, finding a considerably higher peak density of about nine stars per 100 pc$^3$ at $r \sim 15$ pc, and concluded that the density was not necessarily in conflict with the Murray and Sanduleak's result. In my 1985 automated (*m*, *log π*) paper I also analyzed this data (81 stars), similarly finding a peak density of about nine stars per 100 pc$^3$, but at $r \sim 20$ pc as opposed to Dolan's 15 pc. Unfortunately, Dolan did not tabulate his star counts, so we have no way of knowing exactly what were his input data; I have been unable to contact him. Branham also analyzed this data, using 83 stars (see his Table 3), finding a lower peak density of about four stars per 100 pc$^3$ at $r \sim 20$ pc, although followed with a less rapid decline at greater distances than Thé and Staller had found.

I have re-analyzed the Thé and Staller counts as tabulated by Branham using my implementations of the Eddington, Dolan, and Branham methods as well as my 1985 method. The results are shown in Figure 4, along with Thé and Staller's original results. It is not clear to me why my implementation of Dolan's method gives different results from his 1975 paper; my spreadsheet for his method reproduces exactly the results of his Table 1 in his 1974 paper, an analysis of B8-A3 stars in the galactic plane. My implementation of Branham's method also predicts a higher peak density than did he, by about a factor of two. This may be due to different choices of his ridge parameter; my results are for $\tau = 114{,}000$, which gave only one very slight negative density, at $r = 5$ pc. I have contacted Dr. Branham, who was unfortunately unable to recover his notes from the time he was working on this problem and could not recall exactly what ridge parameter or form he used for the Luminosity Function. His estimates of the uncertainties in the derived densities run to about $\pm 50\%$ however, so the discrepancy is likely not as drastic as it appears. I find that if I use a slightly brighter average absolute magnitude (12.0 vs. 12.25) and wider dispersion (0.7 vs. 0.5) with his method, much of the discrepancy between what my implementation of his method returns and his published result disappears, a testament to how sensitive space density analyses can be to the choice of input parameters.

In the end, my implementations of these methods all indicate a significantly higher peak density than that derived by Thé and Staller. Given that the uncertainties with such a small sample size can be significant, their disagreement with Murray and Sanduleak is probably not as great as they believed. That a maximum density occurs at a distance of $\sim 20$ pc is likely a reflection of the fact that this is the approximate displacement of the Sun above the galactic plane (Reed 2006); Thé and Staller did not remark on this point. The controversy generated by their claim appears to have faded from interest; since Branham's 2003 analysis, their paper has been cited only once, in a paper reporting a survey for new late-type low-mass stars.

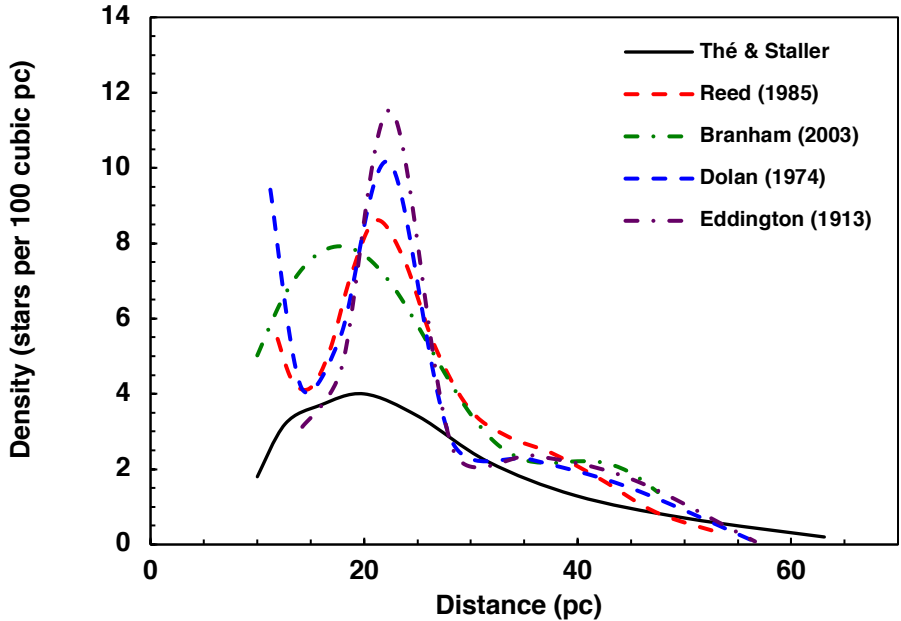

Figure 4. Results of star-count analyses for south galactic pole M-dwarfs from Thé & Staller (1974).

6.2 North Galactic Pole K Giants

Upgren (1962) published an extensive analysis of the distribution of late-type stars in a 396-square-degree region of sky toward the north galactic pole. Like the Thé & Staller M-dwarf counts above, this data makes for a convenient test case as the interstellar extinction is minimal and the fictitious densities can be taken to be real densities. In his 2003 paper, Branham analyzed Upgren's counts for K0 giants, presumably a well-defined spectral grouping. Upgren lists "smoothed" counts of stars per 100 square degrees between $B$ magnitudes 5.0 and 13.0 in half-magnitude intervals; for the K0 giants these total to 155.6 stars, so there must have been some 620 in reality. Upgren performed a traditional ($m$, $log\ \pi$) analysis, assuming a Gaussian LF with $M_O = 1.8$ and $\sigma = 0.8$.

I have re-analyzed these data with the Branham, Dolan, and my 1985 methods; the results are shown in Figure 5. (The Eddington method was tried, but resulted in wild oscillations at nearby distances and so is not included here.) Beyond about 150 pc, my and Dolan's methods track Upgren's results surprisingly closely, while my implementation of Branham's method yields results of about the same magnitude but with the density declining almost linearly down to zero at about 650 pc. (After this distance the Branham density does go slightly negative, but then recovers to slightly positive numbers beyond about 900 pc; this is not shown in Fig. 5.) This again conflicts with what Branham found for this data, and also again we have been unable to pin down the source of this discrepancy. I am confident that my implementation of Branham's method is sound given its solid performance with my simulated data, but in the end remain puzzled at the discrepancies on applying the method to real data. The consistency of results between my method and Dolan's (and, in some cases, Eddington's) in encouraging.

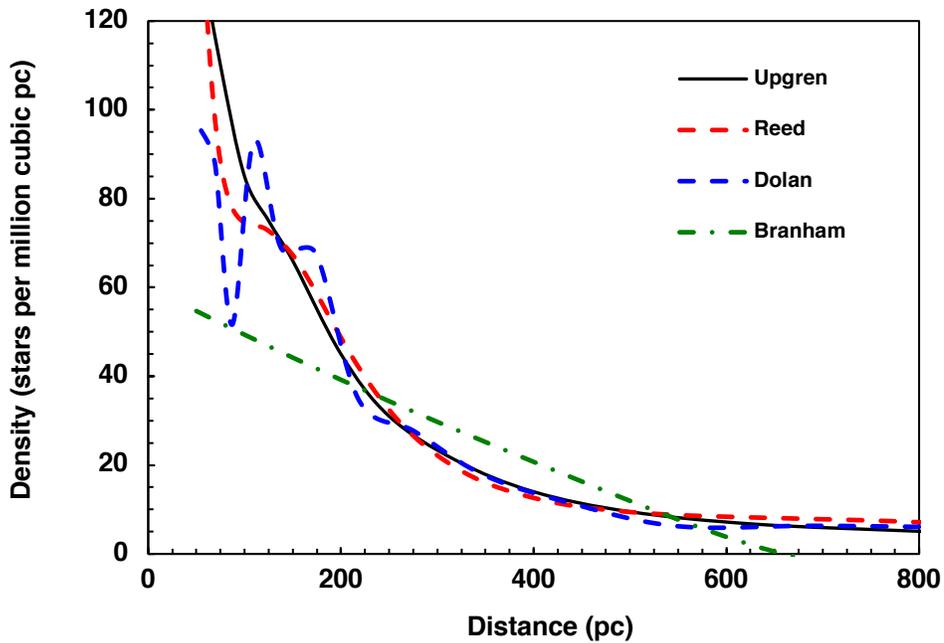

Figure 5. Results of star-count analyses for north galactic pole K giants from Upgren (1962).

## 7. Summary

The main conclusion of this paper is, not surprisingly, that not all methods of analyzing star-count data created equally reliable. The Eddington (1913), Dolan (1974), Reed (1985), and Branham (2003) algorithms are definitely the front-runners. The Reed and Branham methods are preferable in view of the ease with which they could accommodate different luminosity functions. Branham's method is more mathematically sophisticated so far as deconvolving an inherently ill-posed problem is concerned, but possesses the ambiguity of the choice of the ridge parameter, My 1985 method has the advantages of conceptual simplicity and always generating positive densities if the initial guess is positive at all distances, although potentially at the cost of being more prone to the front-loading effect.

Most pre-computer-era space density analyses should probably be regarded with skepticism; an ambitious student of the history of astronomy might wish to compile them all and re-run them in a consistent way using one or more of the methods advocated here. But this is not to say that those studies were meaningless; their fundamental data on thousands of stars are still valid and serve as valuable calibration points for galactic-structure models. Patient human computers were working with the best data, observational technologies, and computational techniques available to them, wringing what they could out of inherently noisy data and no doubt aware of the uncertainties they were up against. In these days when so much data is available at the click of a mouse, we should remember their efforts with respect and humility.

**Acknowledgements**

I am grateful to Richard Branham for discussions on this paper. My interest in star counts goes back to my days as a graduate student in the early 1980's at the University of Waterloo under the mentorship of Dr. Pim FitzGerald, and it is to his lasting good influence that I owe much of my subsequent scientific career.